\documentclass[lettersize,journal]{IEEEtran}
\usepackage{amsmath,amsfonts}
\usepackage{algorithmic}
\usepackage{algorithm}
\usepackage{array}
\usepackage[caption=false,font=normalsize,labelfont=sf,textfont=sf]{subfig}
\usepackage{textcomp}
\usepackage{stfloats}
\usepackage{url}
\usepackage{verbatim}
\usepackage{graphicx}
\usepackage{cite}
\usepackage{booktabs}
\usepackage{tabularx}
\usepackage{caption}
\usepackage{makecell}
\usepackage{color}
\begin{document}

\title{Task-Oriented Wave Processing with Stacked Intelligent Metasurfaces: Framework, Fusion, and Challenges}
%Task-Oriented Wave Processing with Stacked Intelligent Metasurfaces: A Unified Framework for 6G Integrated Services
\author{Qiao Qi, Qiyu Chen, Jiancheng An, Xiaoming Chen, Zhaohui Yang, Chongwen Huang, and Chau Yuen
        % <-this % stops a space
\thanks{
        Qiao Qi and Qiyu Chen and are with the School of Information Science and Technology, Hangzhou Normal University, Hangzhou 311121, China. (e-mail: qiqiao@hznu.edu.cn;cqqqy12138@163.com); 
        Xiaoming Chen, Zhaohui Yang,  and Chongwen Huang are with the College of Information Science and Electronic Engineering, 
        Zhejiang University, Hangzhou 310027, China. (e-mail: chen\_xiaoming@zju.edu.cn; yang\_zhaohui@zju.edu.cn; chongwenhuang@zju.edu.cn); JianCheng An and Chau Yuen are with the School of Electrical and Electronic Engineering, Nanyang Technological University, Singapore 639798. (jiancheng\_an@163.com; chau.yuen@ntu.edu.sg).
        
        \emph{(Corresponding author: Xiaoming Chen)}}
    }

\maketitle

\begin{abstract}
%The deep integration of diverse services, such as communication, sensing, and computation, in sixth-generation (6G) networks poses significant challenges to the conventional task-agnostic wireless channel, often leading to performance conflicts and efficiency bottlenecks. To address this issue, this article introduces a novel stacked intelligent metasurface (SIM)-enabled task-oriented wave processing paradigm, which aims to transform the physical wireless environment from a passive transmission medium into a programmable signal processor. Specifically, we establish a unified framework that systematically maps high-level service requirements to physical wave transformations. We then investigate the fusion of diverse services, demonstrating how the deep computational architecture of SIMs can resolve resource conflicts in integrated sensing and communication (ISAC) and integrated communication and computation (ICC) scenarios. Furthermore, we critically analyze the fundamental physical-layer challenges inherent to this paradigm, including diffractive channel modeling and inverse task-to-phase mapping. Numerical results validate that this approach can effectively enhance synergies in the wave domain, elevating the system from simple coexistence to true service symbiosis. Finally, key future research directions are discussed to pave the way for service-native wireless architectures.

The deep integration of diverse services in sixth-generation (6G) networks poses significant challenges to conventional task-agnostic channels, often resulting in performance conflicts. To resolve these bottlenecks, this article introduces a physical-layer computing paradigm enabled by stacked intelligent metasurfaces (SIMs), transforming the wireless environment from a passive medium into a programmable signal processor. Specifically, we establish a unified framework to map high-level service requirements directly to wave-domain synthesis. We then investigate the fusion of diverse services, demonstrating how the deep computational architecture of SIMs resolves resource conflicts in integrated sensing and communication (ISAC) and integrated communication and computation (ICC) scenarios. Furthermore, we critically analyze fundamental challenges, including diffractive channel modeling and inverse task-to-phase mapping, while validating through numerical results that this approach elevates the system from simple coexistence to true service symbiosis. Finally, we discuss key research directions to pave the way for service-native 6G architectures.
\end{abstract}


\section{Introduction}
The sixth-generation (6G) wireless network is envisioned not merely as an evolution in data transmission rates, but as a revolutionary platform supporting the deep fusion of diverse services, including communication, sensing, computation, artificial intelligence (AI) and so on \cite{Q1,Q2}. Realizing this vision of a seamless service-native ecosystem, however, confronts a fundamental bottleneck rooted in the physical layer. This is because the wireless channel, by its inherent nature, is a passive and task-agnostic medium that indiscriminately propagates and superimposes electromagnetic waves \cite{6G1,6G2}. Consequently, as services with distinct and often conflicting requirements are forced to share this common medium, the channel itself becomes the primary source of interference and performance trade-offs, fundamentally hindering the creation of a true service symbiosis.

To overcome this fundamental impasse, a rethinking of the physical environment's role is necessary, shifting it from a passive hindrance to an active participant in the communication process. This paradigm shift has given rise to programmable wireless environments, with the reconfigurable intelligent surface (RIS) emerging as a prominent technology \cite{RIS1,RIS2,RIS3,RIS4}. Through the digital control of an array of passive elements, an RIS can meticulously manipulate the phase of incident waves to sculpt propagation paths that enhance signal strength and expand coverage \cite{RIS1,RIS2}. Fundamentally, however, the RIS operates as a task-agnostic mirror, a device that enhances the properties of the medium but lacks the inherent capability to process the task-specific information embedded within the wavefronts \cite{RIS3,RIS4}. This critical limitation means that while an RIS can effectively mitigate channel impairments, it is unable to proactively resolve the nuanced conflicts or create the deep synergies required for integrated services, thus falling short of the transformative potential demanded by 6G.

Transcending the limitations of such single-layer surfaces requires a more sophisticated architecture, which has emerged in the form of the stacked intelligent metasurface (SIM) \cite{SIM1,SIM2,SIM3,SIM4,SIM5}. Comprising multiple, densely-packed layers of metasurfaces, the SIM functions less like a planar reflector and more like a physical-domain neural network capable of performing complex signal processing on electromagnetic waves as they propagate through its hierarchical structure \cite{SIM1}. This advanced technological capability underpins the task-oriented wave processing paradigm proposed in this article. The essence of this paradigm is to fundamentally shift the role of the physical layer from a passive medium that must be compensated for, to an active computational resource \cite{SIM2,SIM3}. Instead of reactively compensating for channel effects in the digital domain, this approach proactively engineers the wave propagation itself to perform desired signal transformations \cite{SIM4,SIM5}. In doing so, task-specific objectives are embedded directly into the physical layer. This article leverages this paradigm to establish a unified framework for wave-domain processing, enable the deep fusion of conflicting services, and address the fundamental physical-layer challenges inherent to this evolution. While existing programmable metasurfaces and diffractive neural networks have explored wave manipulation, our framework uniquely integrates a multi-layer deep computational architecture within the wireless context to proactively address multi-objective service requirements, moving beyond single-task optimization or passive channel enhancement. Consequently, the primary contribution of this work lies in establishing this holistic framework, which is specifically tailored to orchestrate complex, integrated 6G services.

%The remainder of this article is organized to systematically formalize the principles of this paradigm and demonstrate its efficacy. Section II establishes the technological foundation by detailing the multi-layer SIM architecture and proposing a unified task-oriented framework that maps service requirements to physical wave transformations. Building on this, Section III investigates the realization of service fusion in typical 6G scenarios and critically analyzes the fundamental physical-layer challenges, such as diffractive modeling and phase optimization. The practical advantages of the proposed framework are numerically validated in Section IV through case studies involving both homogeneous and heterogeneous services. Subsequently, Section V identifies critical future research directions ranging from AI-native configuration to quantum-enhanced hardware architectures. Finally, Section VI concludes the article.

The rest of the article is structured as follows. Section II establishes the technological foundation by detailing the multi-layer SIM architecture and task-oriented framework. Section III analyzes 6G service fusion and related physical-layer challenges. Section IV numerically validates the proposed framework via diverse case studies. Finally, Section V identifies future research directions, such as AI and quantum-enhanced hardware, followed by the conclusion in Section VI.

\begin{table*}[ht]
\centering
\caption{Core differences between RIS and SIM}
\label{tab:ris_vs_sim_transposed_en}
\begin{tabularx}{\textwidth}{ >{\centering\bfseries}m{1.5cm} *{7}{>{\centering\arraybackslash}X} }
\toprule
\textbf{Technology} & \textbf{Core Paradigm} & \textbf{Operational Analogy} & \textbf{Architecture} & \textbf{Primary Function} & \textbf{Capability} & \textbf{Service Handling} & \textbf{System Role} \\
\midrule
\textbf{RIS} & Channel enhancement & Intelligent mirror & Single-layer, \ 2D & Reflects \& Optimize channel state & Computationally shallow & Task-Agnostic Performance trade-off & Channel enhancement component \\
\addlinespace
\textbf{SIM} & {Task-oriented wave processing} & {Physical neural network} & Multi-layer,\ \ \ 3D & {Directly processes \& Transform wavefronts} & {Computationally deep} & {Task-Oriented Service symbiosis} & {Core physical-layer processing unit} \\
\bottomrule
\end{tabularx}
\end{table*}

\section{SIM-Enabled Task-Oriented Wave Processing Framework}

The transition from simple signal reflection to the overarching paradigm of physical-layer computing necessitates a fundamental reimagining of the architecture. While single-layer surfaces offer limited control, achieving complex task-oriented objectives requires a device capable of deep mathematical transformations via multi-layer wave processing. This section delineates the SIM architecture as the hardware foundation for this shift. Based on this platform, we establish a unified framework that maps high-level service requirements into a tailored wave-domain synthesis, enabling the direct execution of computational tasks within the propagation medium.

\subsection{Physical Architecture and Capabilities of SIM}
%SIM represents the evolution from a passive channel enhancer to an active wave processor, enabling a new and powerful paradigm of task-specific signal manipulation. This technology moves beyond the objective of merely improving the channel and instead aims to perform meaningful computational tasks directly on the signal-bearing wave as it propagates. As shown in Fig. 1, the SIM is not a mirror but a lens-like, transmissive structure that transforms an incoming wavefront into a new, highly structured wavefront designed to achieve one or more specific application objectives. This fundamental shift from channel enhancement to in-transit signal processing is what empowers the SIM to actively manage and synergize diverse services, providing the physical-layer foundation for the task-oriented paradigm.
\begin{figure}[t] % [ht] 控制图片位置（h:此处，t:页顶）
    \centering % 图片居中
    \includegraphics[width=0.48\textwidth]{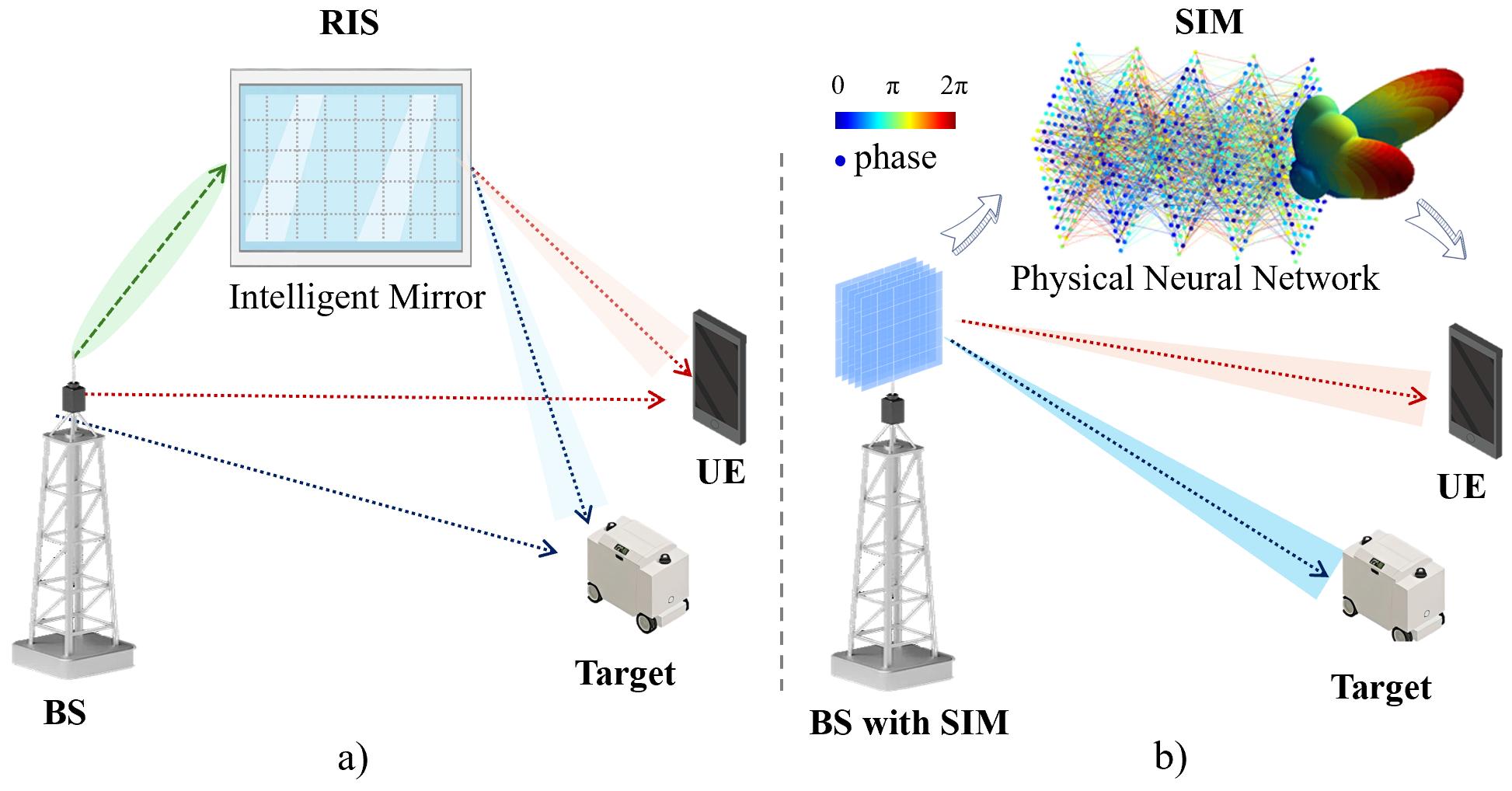} % 宽度设为页面总宽度
    \caption{Architecture comparison between RIS and SIM.} % 图片标题
    \label{fig:ris_sim_comparison} % 标签，用于引用
\end{figure}
Evolving from a passive enhancer to an active wave processor, the SIM executes task-specific signal manipulation directly on electromagnetic waves during propagation. As illustrated in Fig. 1, this lens-like transmissive structure transforms incoming wavefronts into highly structured outputs tailored for specific application objectives. Such in-transit processing establishes the physical-layer foundation for the task-oriented paradigm, enabling the network to actively manage and synergize diverse services through purposeful wave transformations.

\textbf{Physical neural network analogy}:
In stark contrast to the planar RIS, the SIM possesses a computationally deep, three-dimensional (3D) architecture comprising multiple, cascaded layers of metasurfaces, which is functionally analogous to a deep neural network implemented in the physical domain \cite{SIM1,SIM2}. In this analogy, the incident wavefront acts as input data, metasurface layers serve as hidden computational layers, and interlayer diffraction mimics neural connections \cite{SIM3}. While intuitively governed by the Huygens-Fresnel principle, closely spaced SIM layers require advanced electromagnetic modeling to capture near-field coupling and evanescent waves. Despite such physical complexities, this cascaded architecture empowers SIMs to synthesize complex wavefronts unattainable by single-layer structures. Furthermore, unlike purely optical diffractive networks, SIM specifically targets radio frequency (RF) propagation, providing a direct physical-layer computing solution for integrated wireless tasks.

%The architectural depth of the SIM is the critical enabler for its primary capability, i.e., task-oriented wave processing.  This allows the network to move beyond optimizing generic channel metrics and instead embed specific, high-level task objectives directly into the physical wavefront \cite{SIM4,SIM5}.  Unlike simple reflectors that merely redirect energy, the SIM can be configured to execute complex mathematical operations, such as spatial filtering, multi-beam synthesis, or holographic pattern generation, directly on the incident electromagnetic field \cite{SIM-ISAC}. Consequently, the output wave is not merely a steered signal but a synthesized solution where the desired task logic is encoded into the very shape and structure of the propagation.
\textbf{Task-oriented wave processing capability}:
The SIM's architectural depth enables task-oriented wave processing by embedding high-level objectives directly into the physical wavefront \cite{SIM4,SIM5}. Unlike simple reflectors, the SIM executes complex mathematical operations including spatial filtering and multi-beam synthesis directly on incident electromagnetic fields \cite{SIM-ISAC}. The resulting output is a synthesized solution where task-specific logic is encoded into the very structure of the wave propagation. The SIM therefore acts as a physical synthesizer, performing these operations at the speed of light. However, while intuitively promising for offloading power-intensive digital computations, such as high-speed sampling and large-scale matrix inversions, realizing true end-to-end energy efficiency requires carefully balancing these gains against the inherent power overhead of control circuitry and inter-layer insertion losses.

\textbf{Physical-digital co-design}:
Traditionally, signal processing intelligence resides almost exclusively in power-intensive digital basebands. The SIM challenges this by endowing the propagation domain with computational capabilities, enabling a concrete functional split. Specifically, the SIM directly executes wave-domain processing at the speed of light—handling tasks like multi-beam forming, spatial filtering, interference pre-cancellation, and analog computation. Conversely, the digital baseband retains tasks demanding precise arithmetic, such as complex coding/decoding, higher-layer protocols, and network resource management. By strategically offloading these analog-amenable operations to the wave domain, this co-design transforms the physical layer into an active computational stage, significantly relaxing digital hardware constraints.

%The ability to embed tasks into the physical layer precipitates a fundamental shift in wireless system design, forging a new paradigm of physical-digital co-design. Traditionally, the physical layer has been treated as a passive transport medium, with the burden of signal processing intelligence, such as multi-user precoding and interference cancellation, residing almost exclusively in power-intensive digital baseband units. The SIM challenges this rigid division of labor by endowing the propagation domain itself with powerful computational capabilities. This allows for the strategic offloading of complex operations from the digital domain to the wave domain. By performing these tasks via physical propagation, the SIM transforms the physical layer from a passive conduit into an active computational stage, significantly relaxing the performance requirements for digital hardware.

To encapsulate this evolutionary leap from a channel enhancement component to a core physical-layer processor, the fundamental differences between the RIS and SIM paradigms are summarized in Table I.

\begin{figure}[t] % [ht] 控制图片位置（h:此处，t:页顶）
    \centering % 图片居中
    \includegraphics[width=0.45\textwidth]{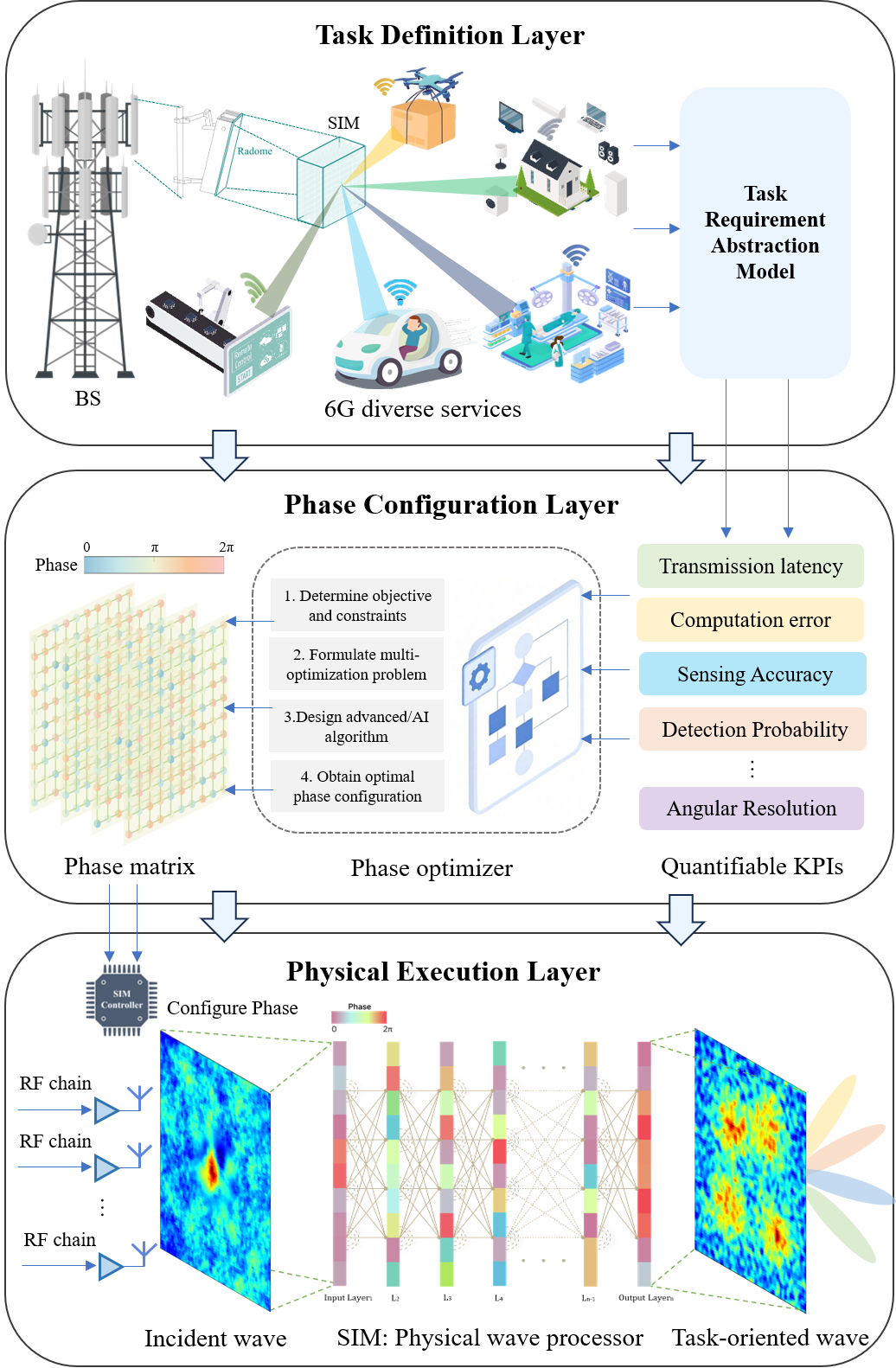} % 宽度设为页面总宽度
    \caption{The three-layer conceptual model of the task-oriented unified framework with SIM.} % 图片标题
    \label{fig:framework} % 标签，用于引用
\end{figure}

\subsection{Task-Oriented Unified Framework with SIM}
To embed high-level service objectives directly into physical wave propagation, we propose a unified three-layer SIM framework, as depicted in Fig. 2. This top-down model translates abstract application requirements into precise electromagnetic transformations, ensuring the entire processing chain is holistically optimized for the target task. The layers are defined as follows:

%The core tenet of the task-oriented wave processing paradigm is to embed high-level service objectives directly into the physical propagation of electromagnetic waves. Based on it, we propose a unified framework with SIM operationalizes the task-oriented paradigm through a three-layer conceptual model, as depicted in Fig. 2. This structured model provides a top-down approach to translate abstract application requirements into precise physical wave transformations, ensuring that the entire signal processing chain, from the digital baseband to the physical wavefront, is holistically optimized for the end-task. Specifically, the definition of each layer is as follows:

\begin{figure*}[ht] % [ht] 控制图片位置（h:此处，t:页顶）
    \centering % 图片居中
    \includegraphics[width=0.85\textwidth]{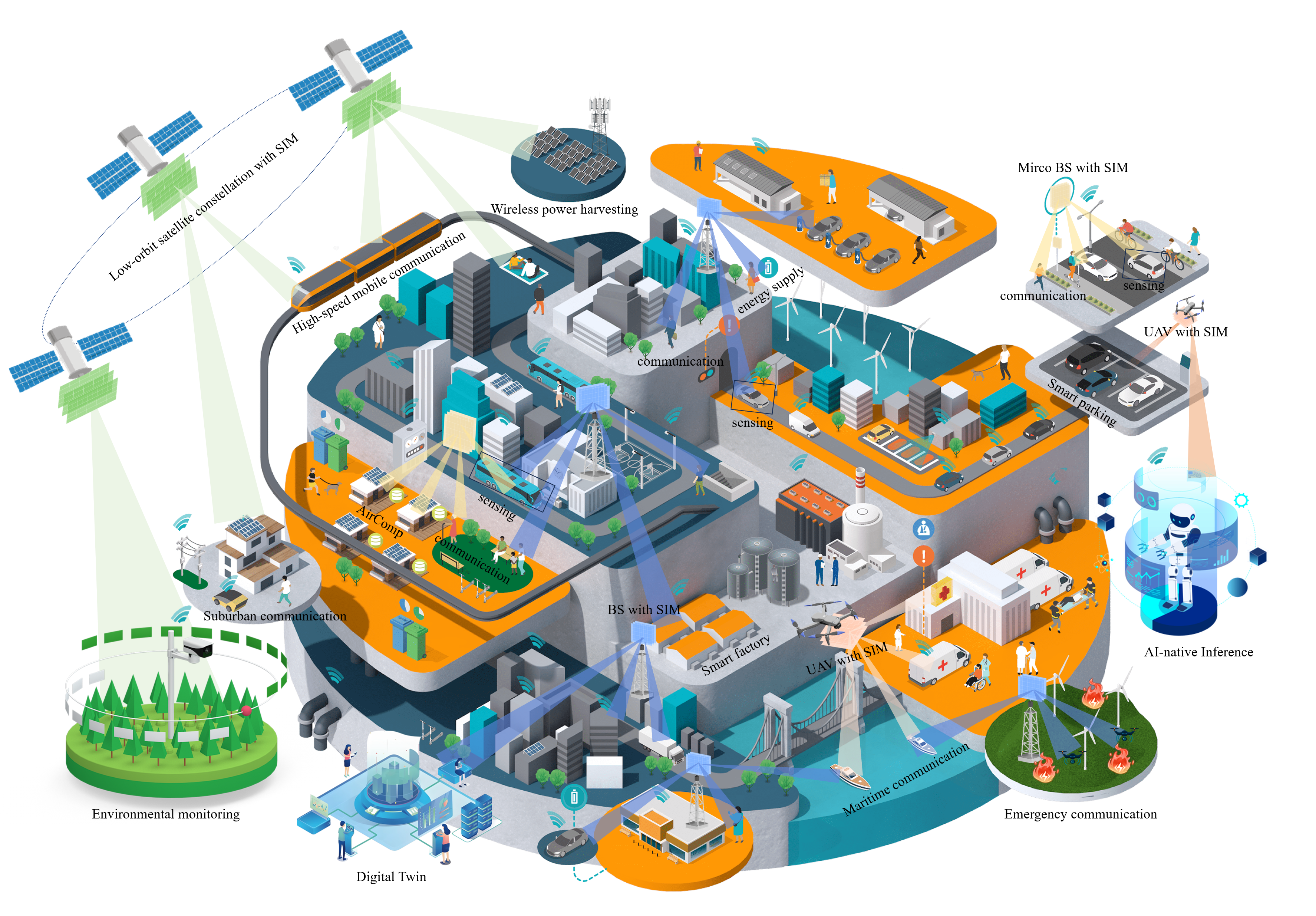} % 宽度设为页面总宽度
    \caption{SIM-enabled 6G service symbiosis scenarios.} % 图片标题
    \label{fig:6Gapplication} % 标签，用于引用
\end{figure*}
\textbf{Task definition layer}: As the highest conceptual interface between 6G applications and the wireless network, it translates service requirements into quantifiable physical-layer objectives. Specifically, a ``task" is defined by a specific set of desired outcomes for a given wireless service, quantified by objective functions or key performance indicators (KPIs) that dictate the desired properties of the electromagnetic wave. For instance, in an integrated service scenario comprising of sensing and communication tasks, this layer would define the required data rate and latency for the communication task, alongside the desired angular resolution and detection accuracy for the sensing task. The success of a task is rigorously quantified by the ability of the SIM to manipulate the wavefront in real-time to meet these specified KPIs, moving beyond generic channel enhancements.

\textbf{Phase configuration layer}: This layer serves as the intelligent bridge between the abstract task and physical hardware, mapping multi-objective KPIs into specific phase shifts. This involves mapping the desired application outcomes onto a specific set of phase shifts that will produce the required electromagnetic wave transformation. Note that the joint phase configuration across a multi-layer SIM is inherently a high-dimensional and non-convex optimization problem, imposing a severe computational burden that restricts the real-time scalability of traditional iterative algorithms. Thus, this layer will likely be powered by advanced algorithms, including AI models, to compute the required phase across all the meta-atoms in the SIM's cascaded structure.

\textbf{Physical execution layer}: This is the physical hardware layer where the SIM resides, including the multi-layered SIM and its control circuitry. It receives the optimized meta-atom phase matrix from the Phase Configuration Layer and implements them by tuning its constituent meta-atoms. The SIM then performs the designed mathematical operation such as multi-beam forming, spatial filtering, or wavefront sculpting on the incident electromagnetic wave at the speed of light. This layer effectively executes a complex algorithm not through digital computation, but through physical wave propagation, achieving unparalleled speed and energy efficiency.

In summary, this framework bridges application intent and physical execution. By vertically integrating task abstraction with intelligent configuration, every phase shift is purposely aligned with the service objective, thereby maximizing spectrum utility and establishing a standard reference for future task-defined networks.
%In summary, this hierarchical framework establishes a seamless pathway from application-layer intent to physical-layer reality. By vertically integrating task abstraction, intelligent configuration, and physical execution, the system ensures that every phase shift within the metasurface is purposefully aligned with the ultimate service objective. This holistic design not only maximizes the utility of the electromagnetic spectrum but also provides a standardized reference model for the implementation of future task-defined wireless networks.

\section{Service Symbiosis: Fusion and Challenges}
Transforming the static physical layer into a programmable substrate is essential for realizing service symbiosis within shared spectrum. By leveraging the SIM's computational depth, resource conflicts are proactively resolved in the wave domain before reaching the digital baseband. As depicted in Fig. \ref{fig:6Gapplication}, this physical-layer processing empowers diverse 6G ecosystems by supporting immersive extended reality (XR) and low-orbit satellite (LEO) constellations through precise spatial interference filtering. Furthermore, it enables high-fidelity digital twins and smart cities via simultaneous environmental sensing and data streaming, while driving industrial Internet of Things (IIoT) and AI-native inference by offloading massive data aggregation directly to the propagation medium.

To demonstrate the versatility of this approach within this broad ecosystem this section investigates two quintessential 6G scenarios. The first is integrated sensing and communication (ISAC) which represents the homogeneous integration of functionality where beamforming conflicts must be resolved within a unified waveform. The second is integrated communication and computation (ICC) which exemplifies heterogeneous integration where the physical layer must simultaneously support data transmission and analog wave-based calculation. The following discussion details how the proposed architecture enables these symbiotic states and subsequently analyzes the fundamental physical layer challenges inherent to such complex integration.

\subsection{Homogeneous Integration via Wave-Domain Functional Splitting}
The integration of sensing capabilities into communication networks traditionally encounters a spatial conflict regarding beamwidth and directionality. High data rate transmission necessitates highly directive beams to maximize signal strength while ubiquitous sensing requires broader radiation patterns to capture environmental information effectively. The SIM addresses this dichotomy by performing functional splitting directly on the electromagnetic wavefront rather than in the digital domain. Through precise diffraction modulation across successive metasurface layers the incident energy is redistributed to simultaneously form high-gain lobes for communication users and independent probing patterns for target detection. This physical synthesis eliminates the need for time-division multiplexing and enables continuous dual-function operations. By shaping the electromagnetic field to serve both purposes concurrently the system transforms the relationship between sensing and communication from one of resource competition to mutually beneficial coexistence.

\subsection{Heterogeneous Integration via Physical Pre-Compensation}
A comparable challenge arises in the ICC where the channel requirements are intrinsically opposed. Over-the-air computation (AirComp) relies on the constructive superposition of signals to perform analog aggregation whereas conventional communication seeks to minimize inter-user interference through orthogonalization. The SIM harmonizes these heterogeneous objectives by functioning as a pre-coding processor within the propagation path itself. The layered structure aligns the phases of computation streams for coherent summation at the receiver while concurrently sculpting nulls to protect parallel communication links. This capability allows the physical medium to differentiate between signal types and execute disparate processing logic simultaneously. Consequently the burden of signal separation and aggregation is strategically offloaded from the digital receiver to the propagation environment.

\subsection{Fundamental Physical-Layer Challenges}
Implementing the task-oriented paradigm necessitates a transition from optimizing simple signal-to-noise ratios (SNR) to manipulating complex wave interference patterns. This fundamental shift from transmission to processing introduces unique theoretical and physical constraints that distinguish the SIM architecture from varying forms of traditional relaying or single-layer reflection.

%\textbf{Inverse task-to-phase mapping complexity}: The primary theoretical hurdle lies in the mathematical intractability of mapping high-level service metrics directly to the physical phase configuration. Conventional digital precoding typically optimizes convex rate expressions where the relationship between the control variables and the output is well-behaved. In contrast the task-oriented paradigm necessitates the optimization of non-differentiable metrics such as sensing detection probability or computation error which exhibit complex dependencies on the electromagnetic parameters. The multi-layer diffraction creates a highly non-linear forward propagation model where the gradient of the task loss function tends to vanish or explode across the cascaded layers. Consequently deriving a closed-form solution for the optimal phase matrix that satisfies the specific quality-of-service requirements for integrated tasks remains a formidable analytical challenge and often precludes the use of standard convex optimization techniques.

\textbf{Inverse task-to-phase mapping complexity}: A primary theoretical hurdle is the mathematical intractability of mapping high-level service metrics to physical phase configurations. Unlike conventional precoding that optimizes well-behaved convex rate expressions, this paradigm targets non-differentiable metrics, e.g., sensing accuracy, computation error, with complex electromagnetic dependencies. Crucially, taking into account the low-level effects of physical layer propagation—such as reactive near-field interactions across sub-wavelength layers—severely exacerbates the overall computational burden. This multi-layer diffraction creates a highly non-linear forward model prone to vanishing or exploding gradients. Consequently, deriving optimal closed-form phase solutions becomes a formidable analytical challenge, often precluding standard convex optimization techniques.

\textbf{Analog computational precision and error propagation}:  Viewed from a signal processing perspective, the SIM functions as an analog optical computer. Unlike error-corrected digital processors, SIMs are susceptible to multiplicative hardware imperfections. Fortunately, multi-layer degrees of freedom relax individual resolution requirements, typically 2 to 3 bits per layer \cite{SIM2,SIM3}, yet near-field diffraction remains sensitive to spatial perturbations. Thus, inter-layer alignment should be strictly maintained within a fraction of the operating wavelength, e.g., $\lambda/10$, to prevent error magnification \cite{SIM1}. Furthermore, reconfiguration latency, ranging from nanosecond-level (e.g., using PIN diodes) to microsecond-level (e.g., using varactors), must be co-designed to match 6G channel coherence. Such phase drifts or timing lags are particularly acute for high-precision tasks like AirComp, where they can destroy the orthogonality required for service fusion, causing inter-task interference unrecoverable by digital basebands.

\textbf{Physical diffraction limits and wave-domain generalization}: The capability of the system to perform functional splitting is bound by the fundamental laws of diffraction and the finite aperture size of the metasurface. The adaptability of the environment is not infinite and there exists a physical limit to the number of distinct beams or independent wave modes that can be synthesized simultaneously. This limit is governed by the degrees of freedom of the stacked array. In scenarios demanding simultaneous wide-beam sensing and pencil-beam communication the system often encounters a resolution bottleneck where the aperture is insufficient to shape the wave with the required spatial sharpness for both tasks. Furthermore the configured wave processor is highly sensitive to the spatial stationarity of the environment. A phase configuration optimized for a specific user distribution acts as a fixed spatial filter meaning that any rapid mobility renders the physical processing logic obsolete and necessitates a re-optimization speed that challenges the physical tuning latency of the hardware.

\section{Case Studies}

In this section, we provide numerical simulations to verify the task-oriented wave processing paradigm with SIM for two typical 6G scenarios, namely ISAC for homogeneous services and ICC for heterogeneous services.  Here, inter-layer propagation within the BS-connected SIM is rigorously modeled via the Rayleigh-Sommerfeld diffraction theory. In practice, this continuous integral is discretized into a layer-to-layer transmission matrix, where each element exactly evaluates the radiative near-field spherical wave propagation, including phase shifts and spatial attenuation, between individual meta-atoms. While capturing fundamental bounds, practical implementations must further account for hardware-specific non-idealities such as cross-polarization leakage, inter-layer misalignment and mutual coupling. Meanwhile, the channel from its outermost layer to the UEs assumes conventional Rayleigh fading to represent small-scale fading effects, ensuring a realistic representation of the wireless environment.

\begin{figure}[t]
    \centering 
    \captionsetup{labelsep=period} 
  
    \captionsetup[subfigure]{
        labelformat=simple,  
        labelfont=rm,     
        font=normalsize     
    }
    \renewcommand{\thesubfigure}{\alph{subfigure})}

    \subfloat[]{\includegraphics[width=0.92\columnwidth]{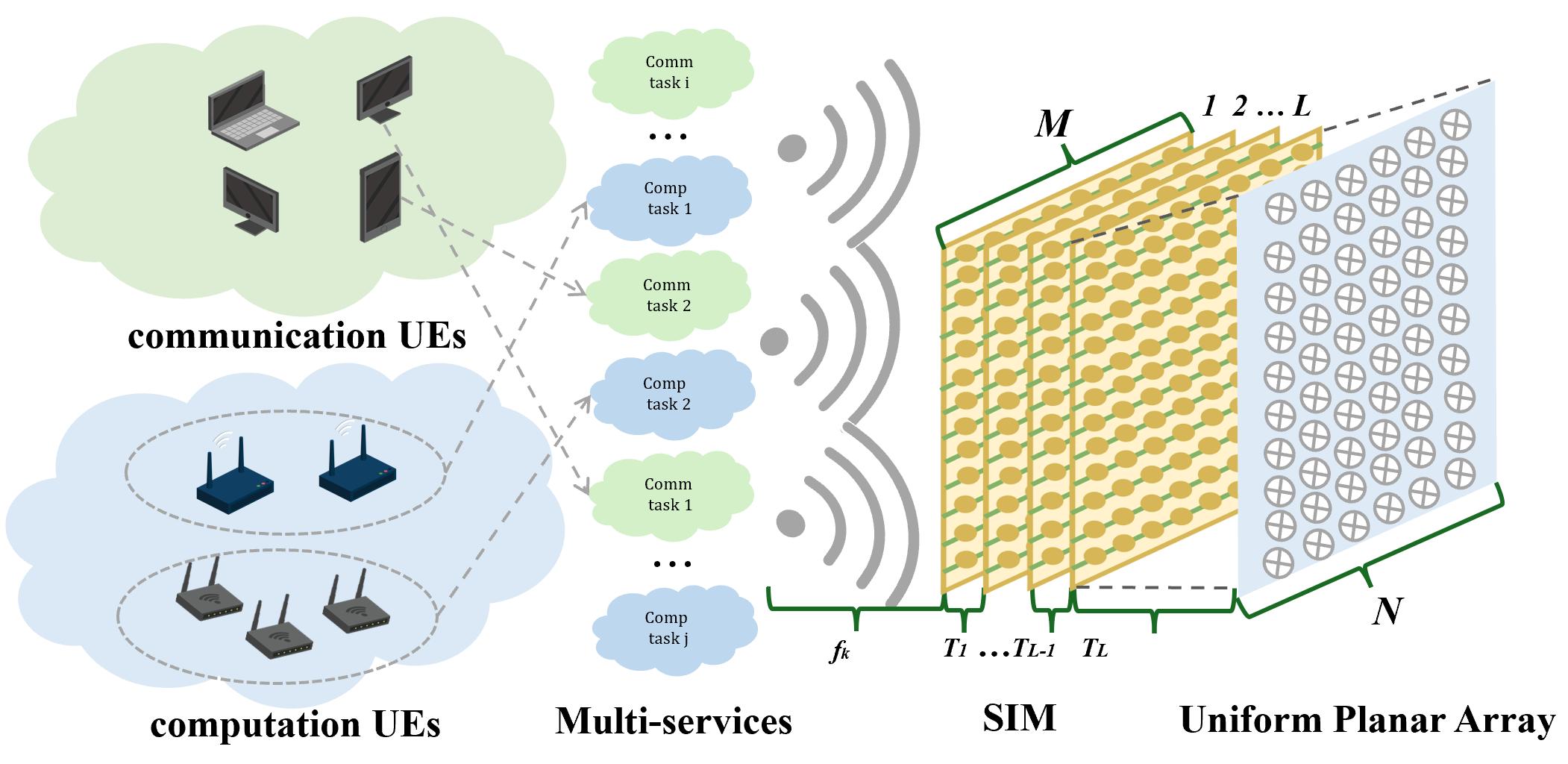}\label{fig:4a}}

    \subfloat[]{\includegraphics[width=\columnwidth]{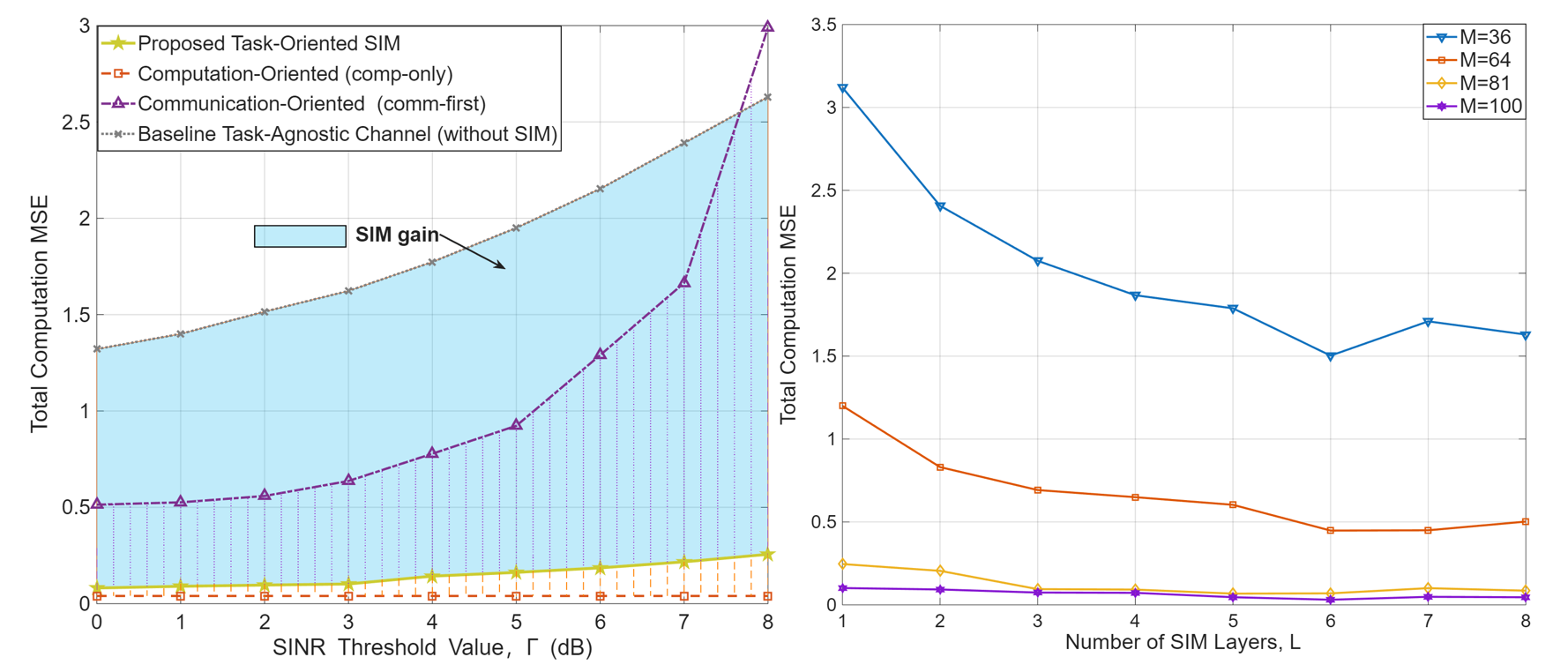}\label{fig:4b}}
    
    \caption{Performance for conflict to coexistence: a) System model of SIM-aided ICC; b) Total computation MSE versus the communication SINR threshold on the left, and performance comparison under different SIM configuration the right (with $K_C=6$ comp UEs and $K_D=3$ comm UEs).}

\end{figure}

\begin{figure}[t]
    \centering 
    \captionsetup{labelsep=period} 
 
    \captionsetup[subfigure]{
        labelformat=simple, 
        labelfont=rm,      
        font=normalsize    
    }

    \renewcommand{\thesubfigure}{\alph{subfigure})}

    \subfloat[]{\includegraphics[width=\columnwidth]{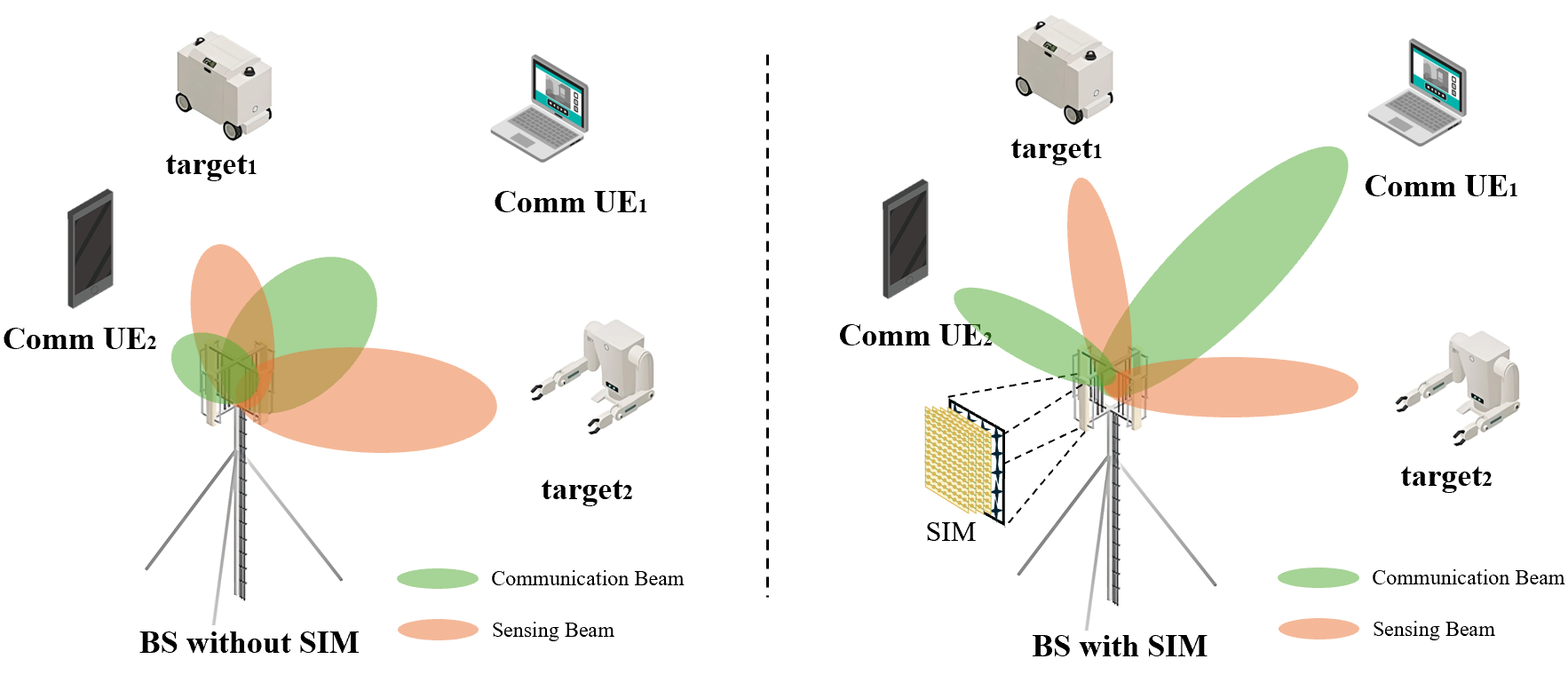}\label{fig:5a}}

    \subfloat[]{\includegraphics[width=\columnwidth]{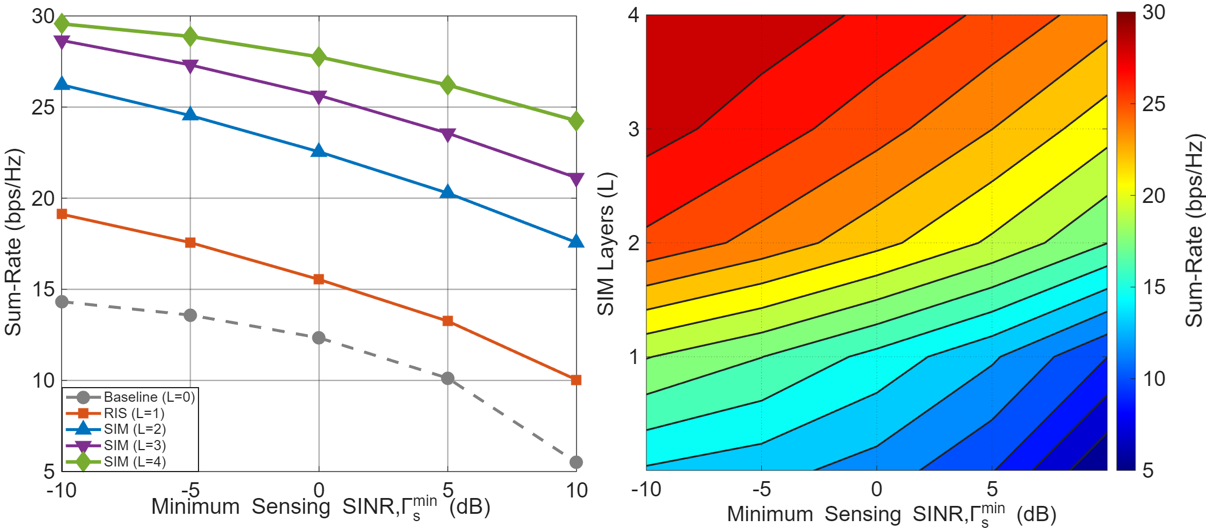}\label{fig:5b}}
    
    \caption{Performance for synergy to symbiosis: a) System model of SIM-aided ISAC; b)  Sum-rate versus sensing SINR threshold and SIM layers on the left, and the corresponding performance contour map on the right (with $K=4$ comm UEs and $S=2$ targets).}

\end{figure}

\subsection{From Conflict to Coexistence for SIM-aided ICC}
We first validate the paradigm in a heterogeneous ICC scenario where a BS equipped with an $L$-layer SIM serves communication user equipments (comm UEs) requiring high signal-to-interference-plus-noise ratio (SINR) and computation UEs (comp UEs) requiring low computation mean square error (MSE) \cite{SIM-ICC}. To demonstrate the superiority of the proposed service fusion framework, we compare it against three benchmarks. The `Task-Agnostic Channel (without SIM)' baseline represents a conventional wireless system relying purely on digital beamforming. Furthermore, we introduce two single-task schemes `Computation-Oriented (comp-only)' and `Communication-Oriented (comm-first)', by allocating all system resource exclusively to one specific task without imposing constraints from the other.

Simulation results in Fig. 4 demonstrate that the task-agnostic baseline fails to manage the interference, while the proposed SIM paradigm achieves a computation MSE comparable to the ideal single-task benchmark and remains robust regardless of the increasing communication SINR threshold. Furthermore, the analysis of computational depth reveals that performance gains saturate at approximately $L=6$ layers, suggesting that excessive depth may lead to wave-domain overfitting. By processing incident waves to create symbiotic virtual channels for both AirComp and communication, the SIM effectively moves the system toward coexistence. Crucially, the inherent wave-domain interference suppression empowers the ICC framework with favorable scalability, ensuring stable performance even as the computation MSE gradually increases with the user count.

\subsection{From Synergy to Symbiosis for SIM-aided ISAC}
To demonstrate the management of homogeneous services, we consider an ISAC scenario where the objective is to maximize communication sum-rate subject to a sensing SINR threshold. Unlike traditional task-agnostic approaches that suffer from beam overlap and inaccurate positioning, the proposed paradigm employs wave-domain functional splitting. The SIM is configured to sculpt the incident wavefront, simultaneously directing high-gain lobes toward comm UEs and customized probing beams toward sensing targets. 

For the ISAC scenario depicted in Fig. 5, we evaluate the system performance across varying numbers of metasurface layers ($L$). In particular, we designate the $L=0$ case as the conventional digital beamforming baseline (without SIM). Furthermore, the $L=1$ configuration specifically models a standard single-layer RIS-aided ISAC system. It is observed that these two baselines suffer severe sum-rate degradation as sensing requirements increase, while the SIM-aided paradigm maintains high performance and robustness. The results indicate that increasing the SIM depth $L$ effectively expands the feasible region of the system, allowing it to sustain high communication rates even under stringent sensing demands. This confirms the ability of the task-oriented framework to physically arbitrate the trade-off, elevating the system from a state of simple synergy to true symbiosis. Furthermore, the proposed architecture exhibits favorable scalability. As the number of comm UEs increases, the SIM leverages its high spatial degrees of freedom to maintain significant multiplexing gains, ensuring robust performance even in high-density user environments.
%Performance evaluations in Fig. 5 contrast the deep computational capability of the multi-layer SIM against a task-agnostic baseline and a single-layer RIS. It is observed that while the baseline and RIS schemes suffer severe sum-rate degradation as sensing requirements increase, the SIM-aided paradigm maintains high performance and robustness. The results indicate that increasing the SIM depth $L$ effectively expands the feasible region of the system, allowing it to sustain high communication rates even under stringent sensing demands. This confirms the ability of the task-oriented framework to physically arbitrate the trade-off, elevating the system from a state of simple synergy to true symbiosis.

\section{Future Research Directions}
Translating this novel task-oriented paradigm into practical deployment faces significant engineering and algorithmic hurdles. To bridge the gap between ideal diffractive models and physical realities, future research must advance algorithmic intelligence, channel acquisition protocols, propagation modeling, and hardware architectures.

%The most immediate priority is overcoming the computational latency associated with the inverse task-to-phase mapping. As discussed in the previous section the optimization landscape for multi-layer diffraction is highly non-convex and precludes the use of standard iterative solvers in dynamic environments. A promising direction lies in the development of AI-native configuration strategies where deep neural networks are trained to approximate the complex inverse function between high-level service KPIs and the low-level phase matrix. Specifically generative models and deep reinforcement learning agents could be employed to infer the optimal hardware configuration in milliseconds by learning the latent features of the wave propagation environment. This approach would effectively shift the computational burden from online optimization to offline training and enable the SIM to adapt to mobile users and changing task requirements with near-zero latency.
\textbf{AI-native real-time configuration}:
The most immediate priority is overcoming the severe computational latency associated with inverse task-to-phase mapping. Since the phase optimization for multi-layer SIMs is inherently high-dimensional and non-convex, traditional iterative algorithms struggle to scale for real-time adaptation in dynamic environments. A long-term conceptual direction to resolve this bottleneck is AI-native configuration. Specifically, generative models or deep reinforcement learning could potentially infer near-optimal phase matrices in milliseconds by shifting the computational burden from online optimization to offline training. However, while theoretically enabling near-zero latency adaptation to user mobility, realizing this ambitious vision faces substantial practical hurdles, primarily the acquisition of massive, high-fidelity electromagnetic datasets required to train these models robustly against unseen propagation conditions.

\textbf{Tensor-based channel reconstruction}:
Acquiring accurate channel state information for a multi-layer diffractive structure remains a formidable bottleneck due to the exponential scaling of pilot overhead. Future efforts should exploit the inherent sparsity of millimeter-wave and terahertz channels by modelling the cascaded transmission tensor through high-dimensional decomposition techniques. By leveraging the low-rank properties of the channel tensor it is possible to reconstruct the full channel matrix from a compressed set of pilot measurements. Furthermore a more radical approach involves bypassing explicit channel estimation entirely through implicit optimization. In this scheme the system uses end-to-end learning to optimize the beamforming vectors and phase shifts directly based on task feedback such as bit error rate or sensing detection accuracy rather than reconstructing the intermediate channel matrix.

\textbf{Near-field exploitation and wideband management}:
As 6G systems migrate towards extremely large apertures and terahertz frequencies the fundamental propagation conditions shift from the far-field planar wave assumption to the radiative near-field regime characterized by spherical wavefronts. This transition presents a unique opportunity to exploit the distance dimension for beam focusing where energy is concentrated at a specific 3D spatial point rather than along an angular direction. Future research must explore how SIMs can manipulate these spherical waves to create spot-beams for enhanced physical layer security and spatial multiplexing. Concurrently the inherent frequency-dependent diffraction of metasurfaces introduces severe beam squint in wideband transmission. Developing frequency-resilient phase modulation techniques or designing dispersive meta-atom structures that can maintain stable beam alignment across ultra-wide bandwidths constitutes a critical physical-layer imperative for practical deployment.

\textbf{Active and quantum-enhanced architectures}:
While the current framework focuses on passive metasurfaces to minimize energy consumption the inherent insertion loss of multi-layer structures imposes a limit on the achievable end-to-end gain. The integration of active reflection amplifiers into the metasurface layers represents a critical hardware evolution. By enabling simultaneous phase and amplitude variations, an active stacked architecture would not only compensate for diffractive losses but also allow for non-linear wave processing capabilities such as frequency mixing and signal amplification directly in the wave domain. Looking further ahead, the intersection of metasurface technology and quantum information science remains a highly speculative, long-term theoretical frontier. While manipulating quantum states, e.g., entanglement, could hypothetically enable sensing beyond classical diffraction limits, translating these quantum-optical concepts to microwave SIMs requires overcoming currently prohibitive hardware complexities, material limitations, and severe decoherence effects in dynamic wireless channels. %Looking further ahead the intersection of metasurface technology and quantum information science offers a visionary frontier. Quantum-enhanced SIMs could theoretically manipulate quantum states of light such as entanglement and squeezing to achieve sensing resolutions beyond the classical diffraction limit or to enable unconditionally secure quantum key distribution channels embedded within the physical wave processing layer.

\section{Conclusion}
%This article establishes a task-oriented wave processing paradigm that redefines the physical layer from a passive medium into a programmable computational resource. Leveraging the deep architecture of SIMs, we proposed a unified framework embedding high-level service objectives directly into electromagnetic wave propagation. Numerical validations demonstrated that this approach effectively resolves 6G resource conflicts in both homogeneous ISAC and heterogeneous ICC scenarios. Ultimately, by shifting from digital mitigation to wave-domain synthesis, this work positions the SIM not merely as an auxiliary channel enhancer, but as a fundamental processing unit essential for achieving genuine service symbiosis in future wireless architectures.

This article establishes a task-oriented wave processing paradigm that transforms the wireless environment into a programmable computational resource. By leveraging the deep architecture of SIMs, our proposed framework embeds high-level service objectives directly into electromagnetic wave propagation, effectively resolving 6G resource conflicts in both ISAC and ICC scenarios. Shifting from digital mitigation to wave-domain synthesis positions the SIM as a fundamental processing unit, paving the way for genuine service symbiosis in future 6G architectures.
%This article has established the theoretical and practical foundations for task-oriented wave processing, a paradigm that redefines the physical layer from a passive transmission medium into a programmable computational resource. By utilizing the deep architecture of SIMs, we developed a unified framework capable of embedding high-level service objectives directly into the propagation of electromagnetic waves. Numerical validations in representative heterogeneous and homogeneous scenarios confirmed that this approach effectively resolves the inherent resource conflicts typical of 6G networks. The transition from digital-domain mitigation to wave-domain synthesis was shown to elevate system performance from mere coexistence to genuine service symbiosis. Ultimately, this work positions the SIM not as an auxiliary channel enhancement component but as a fundamental processing unit essential for the intelligent, service-native evolution of future wireless architectures.

\section*{Acknowledgments}
%This research is supported by xxxx

The work is supported in part by the Natural Science Foundation of China under Grant 62501222, 62231009, 62331023 and 62394292, Zhejiang Provincial Natural Science Foundation of China under Grant LQN25F010014, Zhejiang Key R\&D Program under Grant 2023C01021, Young Elite Scientists Sponsorship Program by CAST 2023QNRC001, Fundamental Research Funds for the Central Universities under Grant K2023QA0AL02, and Ministry of Education, Singapore, under its MOE Tier 2 (Award number T2EP50124-0032).

\section*{Biographies} % 加星号去掉罗马数字序号

% 开启小号字体环境。如果你觉得 \small 还是不够小，可以把 \small 替换为 \footnotesize
\begin{footnotesize} 

\noindent \textbf{Qiao Qi} (qiqiao@hznu.edu.cn) is a Lecturer with the School of Information Science and Technology, Hangzhou Normal University (HZNU). She received the PhD degree from Zhejiang University (ZJU) in 2023. Her research interests include cellular IoT, edge intelligence, and integrated sensing, communication and computing (ISCC).

\vspace{0.8em} % 因为字号变小了，段落之间的间距也顺势从 1.5em 调小到 1em，显得更紧凑

\noindent \textbf{Qiyu Chen} (cqqqy12138@163.com) is currently pursuing a B.S. degree with the School of Information Science and Technology, HZNU. His research interests include intelligent communication and stacked intelligent metasurfaces (SIM).

\vspace{0.8em}

\noindent \textbf{Jiancheng An} (jiancheng\_an@163.com) is a Research Fellow with the School of Electrical and Electronics Engineering (EEE), Nanyang Technological University (NTU). He received the Ph.D. degree from the University of Electronic Science and Technology of China (UESTC) in 2021. His research interests include wireless communication, SIM and wave-based computing.

\vspace{0.8em}

\noindent \textbf{Xiaoming Chen} (chen\_xiaoming@zju.edu.cn) is a Professor with the College of Information Science and Electronic Engineering (ISEE), ZJU. He received the Ph.D. degree from ZJU in 2011. His research interests include 5G/6G key techniques, satellite communication,  and smart communications.

\vspace{0.8em}

\noindent \textbf{Zhaohui Yang} (yang\_zhaohui@zju.edu.cn) is a ZJU Young Professor with the College of ISEE, ZJU. He received the Ph.D. degree from Southeast University in 2018. His research interests include ISCC, federated learning, and semantic communication.

\vspace{0.8em}

\noindent \textbf{Chongwen Huang} (chongwenhuang@zju.edu.cn) is a tenure-track young professor with the College of ISEE, ZJU. He received the Ph.D. degree from Singapore University of Technology and Design (SUTD) in 2019. His research interests include Holographic MIMO Surface, B5G/6G Wireless Communications, mmWave/THz Communications.

\vspace{0.8em}

\noindent \textbf{Chau Yuen} (chau.yuen@ntu.edu.sg) is an Associate Professor with the School of EEE, NTU. He received the Ph.D. degree from NTU in 2004. He is a Distinguished Lecturer of IEEE Vehicular Technology Society, Top 2\% Scientists by Stanford University, and also a Highly Cited Researcher by Clarivate Web of Science.

\end{footnotesize} % 结束小号字体环境

\vfill

\end{document}